\documentclass[12pt]{revtex4}
\usepackage{epsfig}
\begin{document}
\pagestyle{empty}

\title{
 \hskip 3.0in{\normalsize \bf PREPRINT ITEP-07-09}\\
\bigskip
\bigskip
{\bf Dense Cold Nuclear Matter Study \\
with Cumulative Trigger.\\
 Proposal}
} 
\author{
O.~Denisovskaya$^a,^b$, K.~Mikhailov$^a$, P.~Polozov$^a$,
M.~Prokudin$^a$, G.~Sharkov$^a$, A.~Stavinsky$^a$,
V.~Stolin$^a$, R.~Tolochek$^a,^b$, S.~Tolstoukhov$^a$
}
\affiliation{$^a$Institute of Theoretical and Experimental Physics,
B.~Cheremushkinskaya 25, 117259, Moscow, Russia\\
$^b$National Research Nuclear University "MEPhI",
Kashirskoe Shosse 31, 115409 Moscow, Russia
}

\begin{abstract}
Experimental program for the study of dense cold matter is proposed. 
Droplets of such  a matter are expected to be created in light ion collisions 
at the initial energy range of future facilities FAIR and NICA with extremely 
small but measurable cross section. Meson (or photon) production at high $p_t$ 
and central rapidity region (double cumulative processes domain) is proposed 
as possible effective trigger (selection criteria) for such study.
\end{abstract}

\maketitle

\newpage
\pagestyle{plain}
\pagestyle{empty} 

\section*{\vskip 2.0cm \LARGE 1. Introduction}
\begin{figure}[!hb]
\centering
\includegraphics[width=16cm]{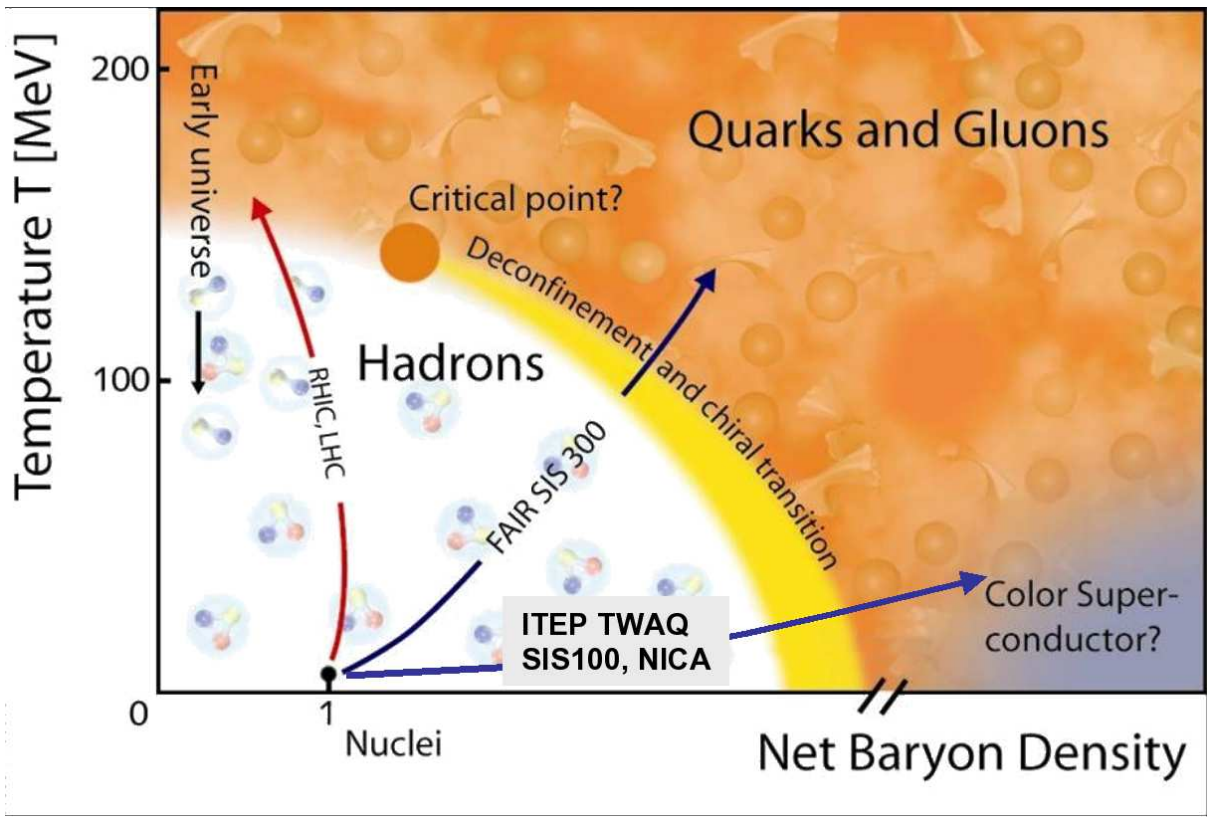}
\caption{Phase diagram of nuclear matter.}
\label{fig1}
\end{figure}

\hskip 0.7cm Chromodynamics  of media is  the subject  of research  in relativistic
nuclear  physics field.   From theoretical  point of  view it  is very
important to study  the phase diagram structure in  QCD in detail (see
\textit{e.g.}   \cite{1}, Fig.\ref{fig1})  and to  find specific  signatures in
nuclear interactions indicating on the phase transitions presence. One
of the main goals for experiments  on heavy ion beams is discovery and
study of new  form of QCD matter --  quark-gluon plasma (QGP) \cite{2}.
At present  the main experimentalists' efforts are  bended on studying
the phase diagram at high temperatures and low baryon densities (RHIC,
LHC) \cite{3}.   This corresponds to  the theory status ten  years ago
when the phase diagram consisted only of two regions: hadron phase and
QGP.  Since resently the progress in the theory led  to the  significant
complication  of the phase diagram. 
In particular, it has led to the  appearance of
critical point~\cite{4}.  Discovery  of critical point at intermediate
temperatures and densities is considered  as one of the most important
goals of FAIR and NICA  projects.  New phenomena are also predicted at
high densities and low temperatures.  In this phase space domain the first
order  phase transition  and  the existence of  new  phenomena like  color
superconductivity~\cite{5}  is expected.  Low temperatures  and extreme
densities are probably realized  in the Nature within neutron stars.  This
region  is hardly   achieved in  laboratory conditions  by using
standard experimental tools 
like changing
of the initial energy and masses of colliding nuclei, or selecting the
impact parameters.  Such tools don't provide possibility  to study the
whole phase diagram, but they  specify some rather small area $T \pm dT$ versus
$\rho \pm d \rho$.

Cumulative  effect which has  been discovered  in 1970's  \cite{6} has
been considered in terms of dense fluctuations of nuclear matter. Some
properties of  cumulative processes, such  as strangeness enhancement,
are  similar to  that expected  for QGP.   However  interconnection of
cumulative  processes and  QGP seems  to be  doubtful because  of next
arguments. Firstly, if high  densities could be realized in cumulative
processes,   then   only  in   short-lived   fluctuations  (named   by
D.I.Blohintsev  ``fluctons'' \cite{7}) could be created.  Secondly,  particles in  such
fluctuations  must   be  highly   virtual  and  they have   large  relative
momentum. Thirdly, these (fluctons) are  local few-nucleon fluctuations and it is
inconsistently  to consider  them like  a media  (although  existence of
plasma  droplets was  already  discussed, see  for example~\cite{8}).
Recognizing  significance   of  these  objections,
we propose an
effective trigger for extreme dense nuclear matter. Indeed if we could
select (\textit{e.g.}  kinematically) a process where approximately ten nucleons being
in volume of one nucleon, then  the density of such formation would be
tens  times higher  the standard  nuclear density  (0.17 nucleons/fm$^3$).
Our proposed  program  gives  an  idea  how to  overcome  mentioned  
problems above.

\section*{\hspace  {5  mm} \LARGE 2. General idea} \hspace  {5  mm}  

\begin{figure}[!ht]
\centering
\includegraphics[width=0.9\textwidth]{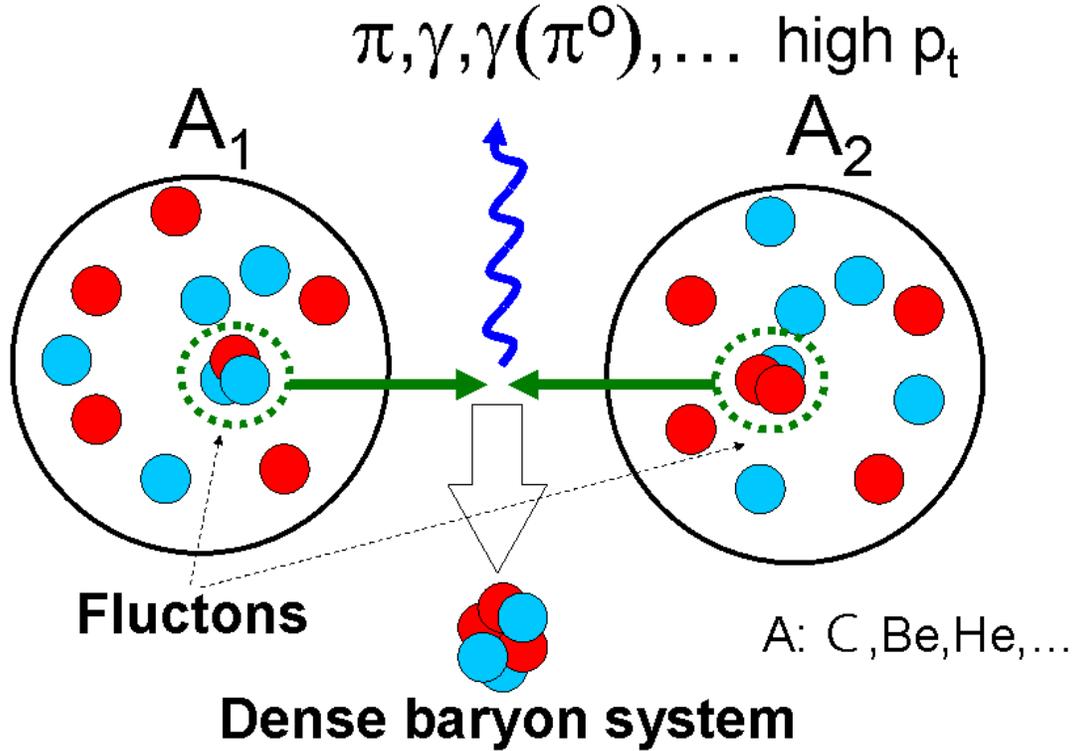}
\caption{ Scheme of the process.}
\label{fig2}
\end{figure}

\begin{figure}[!ht]
\centering
\includegraphics[width=1.0\textwidth]{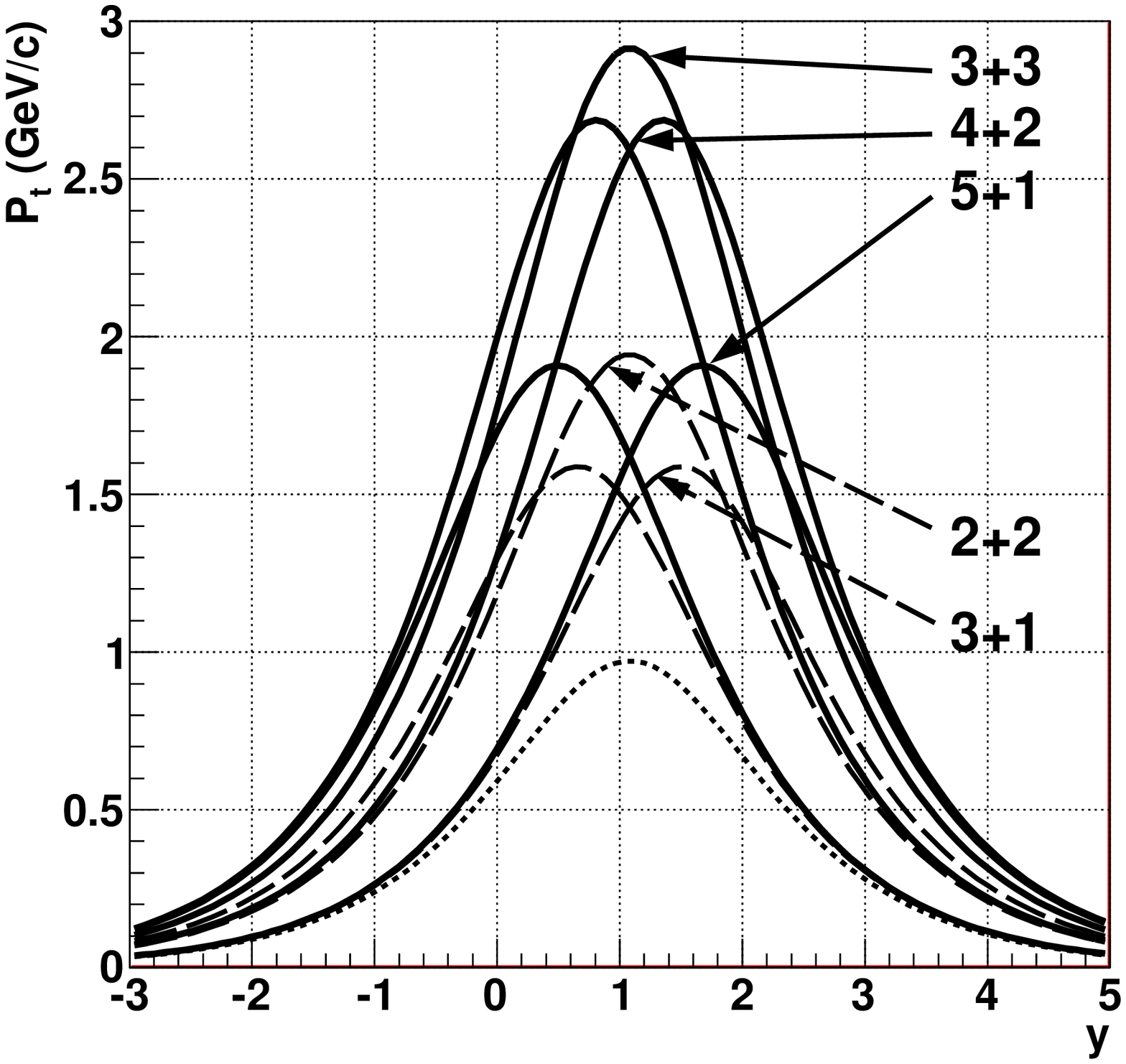}
\caption{Kinematical limits for iN+jN$\rightarrow \pi^0$ +X ($i+j \le 8$) 
processes at beam energy 6AGeV. Curves: i+j=2 (1N+1N) - dotted line; i+j=4 (3N+1N, 1N+3N and 2N+2N) - dashed lines;
 and i+j=6 (5N+1N, 1N+5N, 4N+2N, 2N+4N, 3N+3N) - solid lines.}
\label{fig3}
\end{figure}

We  propose to  make an event  selection (trigger)  with a  photon (pion,
kaon)  at mid-rapidity  and  maximal transverse  momentum by  colliding
light  nuclei  (from  Helium   to  Carbon)  (Fig.\ref{fig2}).  Due  to
kinematical restrictions such criterion selects mainly flucton-flucton
(FF) interaction (Fig.\ref{fig3}). We should stress that production of
cumulative particle is neither  necessary nor sufficient condition for
selection  of  dense  baryon  system.  However, we  expect  that  such
selection   procedure  would  increase   signal  (dense   cold  matter
production) to background (ordinary hadronic matter) ratio for several
orders of magnitude.  
The interaction of two fluctons produces a system (firewall) with real particles (nucleons).  
The internal energy of the secondary baryon system (recoil system) becomes 
minimal when the energy of the trigger particle tends to the maximum 
possible energy for present colliding nuclei reaction.
Therefore the decay of this system (with very small internal energy) will be slow. 
The relatively slow decay of this system could restrict sort lifetime 
of this system and large relative momentum of secondary baryons.
Thereby  such a system should have high density and small size, and one can
speak about medium, since the free path will be much smaller then the size
of the system due to high density.

After realization  of proposed event  selection we suggest  to proceed
with  a bright  research program  focused on  properties of  formed   system in
final  state.   Theoretically  predicted properties  of  dense
baryon  system (some  of  them  are listed  below)  should be  checked
experimentally. This list will probably become longer in future, but it
seems  already  clear  nowadays  that spin(isospin)  system's  states,
space-time characteristics, search of exotic particles (such as dibaryons),
strangeness enhancement etc. should be studied.  In this context it is
important  to mention  research  programs discussed  in the mid 80's  and
related to cumulative process studies~\cite{9}. 
Some of these proposals (though with modifications) and a list of considerations are an important issue today.

\begin{figure}[!hb]
\centering
\includegraphics[width=1.0\textwidth]{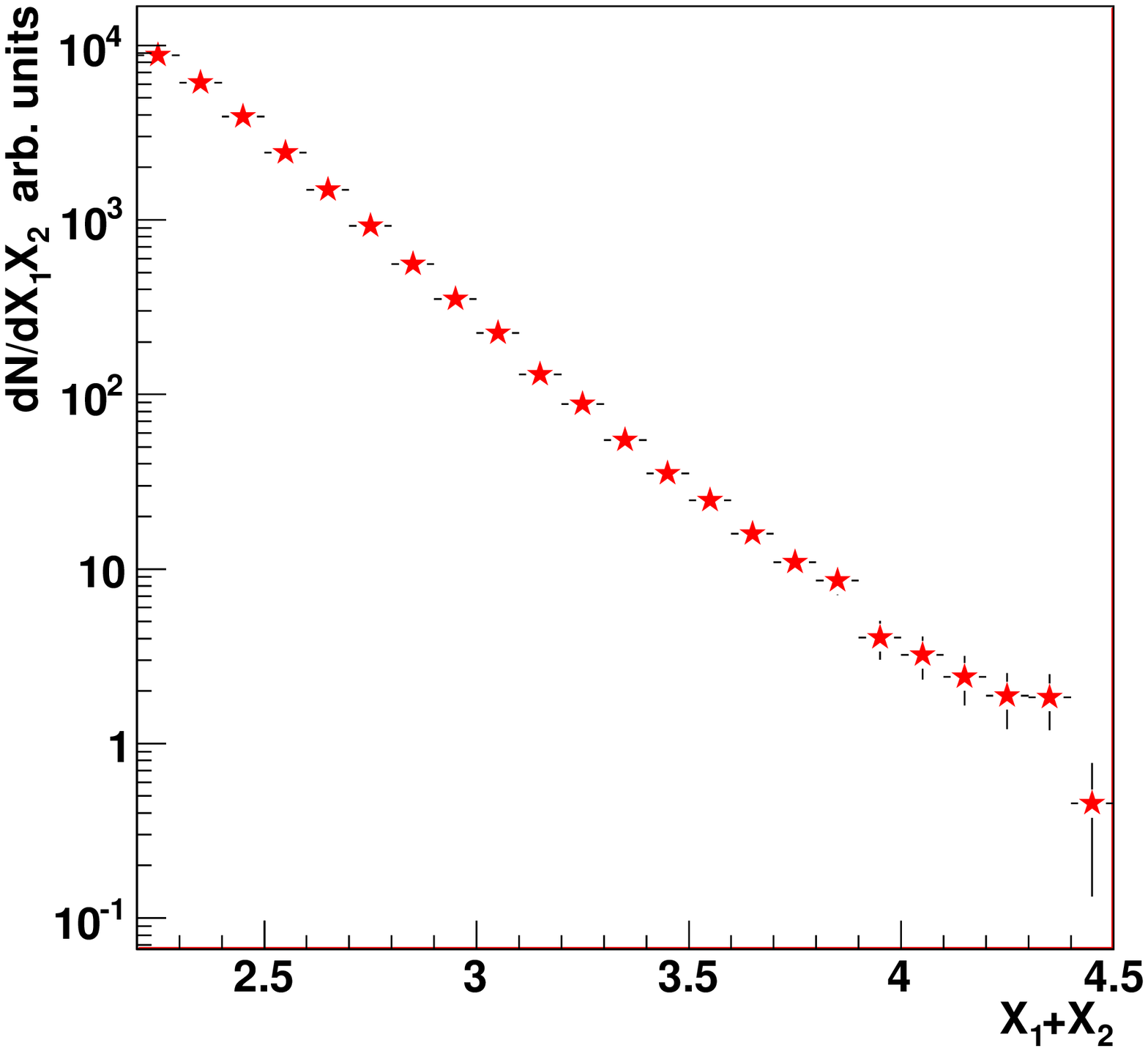}
\caption{Photon spectrum measured by FLINT collaboration for Be(3.2AGeV)C 
interaction vs photon cumulative number.}
\label{fig5}
\end{figure}

Reality of  an effective trigger for the  selection of flucton-flucton
interaction  was experimentally  proven in  FLINT (FLuctonINTeraction)
experiment  in ITEP~\cite{10}. The  trigger realization  in  FLINT was
based  on high $p_t$  photon registration  in mid-rapidity  range using
lead-glass electromagnetic  calorimeter. Type  of the  calorimeter is
uncritical, but  its energy resolution should be  good enough (~100-150
MeV) for  cumulative photon  spectra
measurements. Maximal  order of
cumulativeness  $X_{sum} =  X_1+X_2$ (where  $  X_i $  is the  minimal
number of participating nucleons from colliding nucleus $i$=1,2) achieved at
the first  stage of the  experiment was about $X_{sum} = 5$ (Fig.\ref{fig5}). 
The cross section of the reaction decreased by $2 \div 3$ order of magnitude
when   $X_{sum}$ increased by one unit.
The value $X_{sum} \sim 7-8$ could
be accessible experimentally with large acceptance detector at ion-ion
interaction rate  $\sim 10^8 sec^{-1}$.  
The nature  of the triggered photon  is not  important for  the study.
It could be direct photon or the photon from  unstable particle decay
(mainly from $\pi^{0}$). The kinematical limit (or $X_{sum}$) of photon production
is very close pion one.
Pion, kaon or even $\phi$-meson trigger is also  possible from physical point of view,
and the possibility of practical realization should be discussed.

\section*{\hspace  {5  mm}\LARGE 3. Specific features of  the proposed  program} \hspace  {5 mm}

	Proposed  trigger is  only  the  tool to  move  us into  phase
diagram  domain   not  studied  yet.   
One  needs   to  realize the breadth of the
experimental  program to study the  dense  baryonic system  
which is expected to be produced in  the selected events.  
Each signature of the
dense  cold matter  should  be checked  experimentally  and the  total
experimental  program can  be as  wide  as the existing  program for  
the high temperature plasma study (RHIC,  LHC, FAIR). 
The detector should be adequate to different tasks which are proposed 
to study various  properties of the dense baryon system. 
This requirement is specific for each task and will be discussed below separately.

\subsection*{\hspace  {5  mm}\LARGE 3.1. Clusterization}

\begin{figure}[hbt]
\centering
\includegraphics[width=0.49\textwidth]{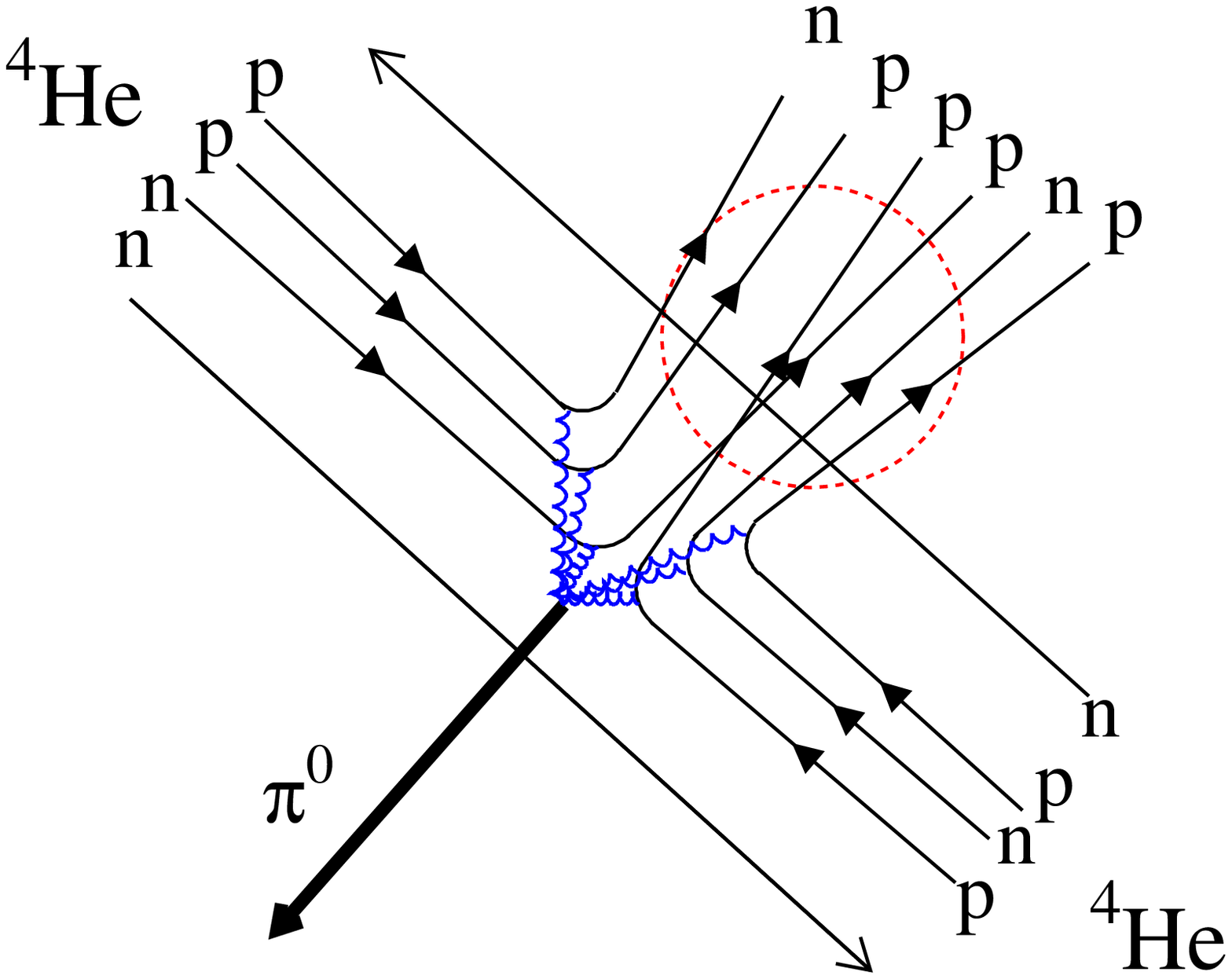}
\includegraphics[width=0.49\textwidth]{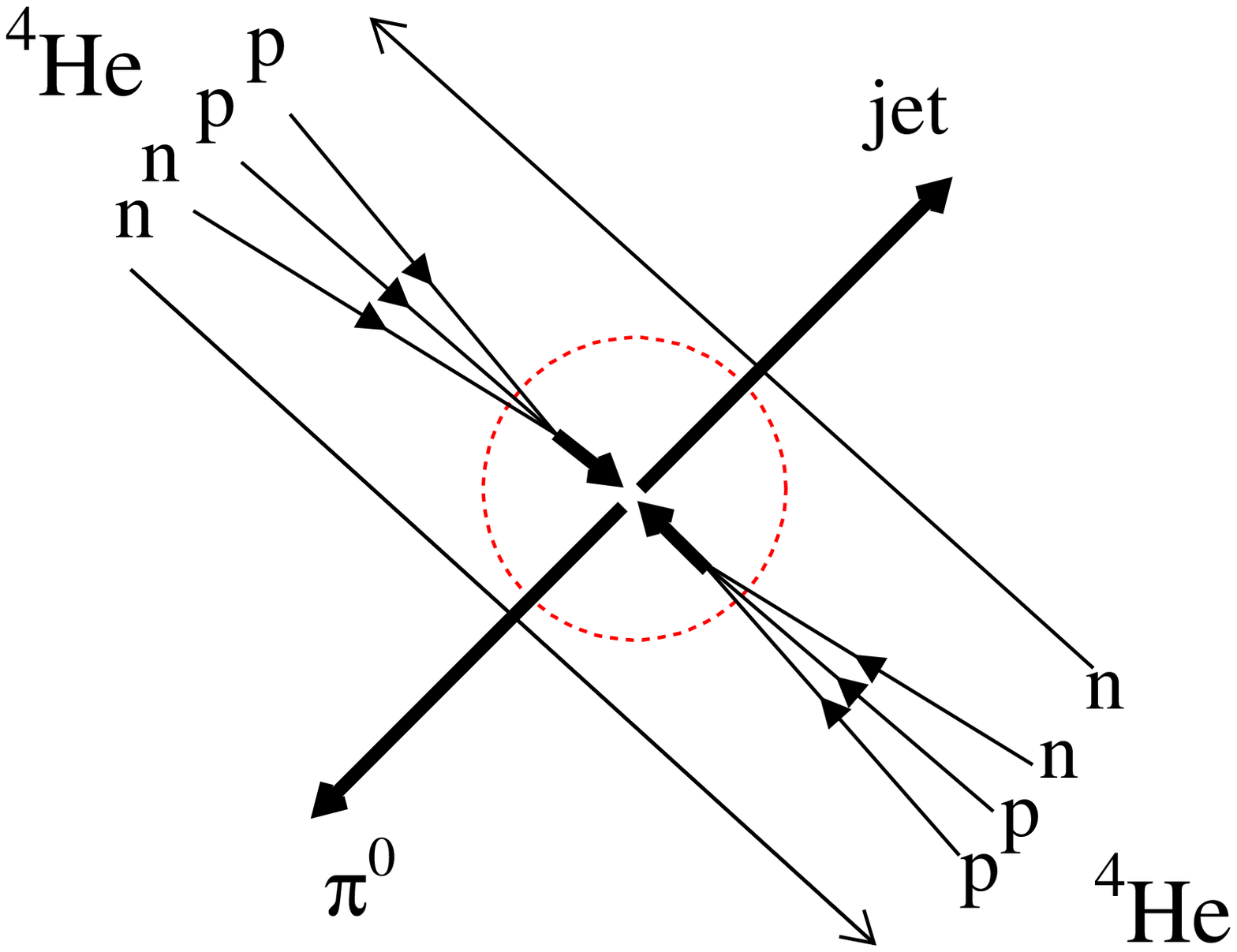}
\caption{Possible scenario for the cumulative process  
$3N+3N \rightarrow \pi^0 + X$ at  HeHe interaction. }
\label{fig6}
\end{figure}

\hskip 0.7cm 	One  of  the  expected  consequences  of  the  proposed  event
selection  is baryon clusterization  in momentum  space in  the final
state. We’ll  refer below phase  space region where  clusterization is
expected as  recoil baryon rich bubble region (B2R3). The position of the B2R3 
in the momentum space  is model  dependent (Fig.\ref{fig6}). In  quasi-binary
model   (see Fig.~\ref{fig6}left)  of  flucton-flucton   collision  cluster
momentum could be defined as 
$P_c \sim P_{f_1}+P_{f_2}-P_{trig}$,
where $P_{f_i}$ ($i$=1,2) is the momentum of flucton, 
and $ P_{trig}$ is the momentum of the triggered particle.

The cluster momentum is close zero (in c.m. system of colliding nuclei) 
in the model where one of  the partons 
(quarks) caries most of the flucton momentum  and triggered  
particle is produced  in the hard interaction  of two partons
(see Fig.~\ref{fig6}right).

It is  expected that the B2R3 will be  at mid-rapidity region in both model,
but the momentum of cluster will be different. The other models which describe 
two nucleus collision  are possible,  
but we restrict our considerations  with these models.
One can expect much more narrow relative momentum distribution (in final state)
of baryons which were involved  in the selected flucton-flucton interaction
in comparison with relative momentum distribution of other baryons (spectators)
 produced in light ion collision.
For  heavier  than HeHe  colliding  system 
(\textit{e.g.}   CC),  one can  expect  a  gap  in baryon-relative-momentum
distribution. This gap separates baryon pairs participating in dense
and  cold baryon  system  from other  pairs.  Spectators of  colliding
nuclei (which clusterization is trivial) should be excluded from
analysis. Thus  to see the  expected clusterization effect  one should
identify and measure  baryon momenta at mid-rapidity region ($ y  \sim 0 $) and 
transverse momentum region upto momentum of triggered particle ($ 0  < P_t < P_{trig}  $).  
The momentum measurement accuracy should be  better than typical
Fermi momentum ($\Delta p \lesssim P_F$).

\subsection*{\hspace  {5  mm}\LARGE 3.2. Femtoscopy}

\hskip 0.7cm One can expect  also a clusterisation of baryons (which was  involved in selected
flucton-flucton  interactions)  in the  coordinate  space and  the increase  of
cluster throwout  time in comparison  with throwout time  of particles
which  are  not  in  the  same cluster.  These  predictions  could  be
experimentally       tested       using       correlation       method
(femtoscopy). The femtoscopy method based on the pair correlation function
analysis  at  small  relative  momenta for  different  particle  types~\cite{11}.
Correlation functions can  give information about size and
possibly  about form of  source.  The  method is  widely used  in  heavy ion
collisions.

\begin{figure}[hbt]
\centering
\includegraphics[width=1.0\textwidth]{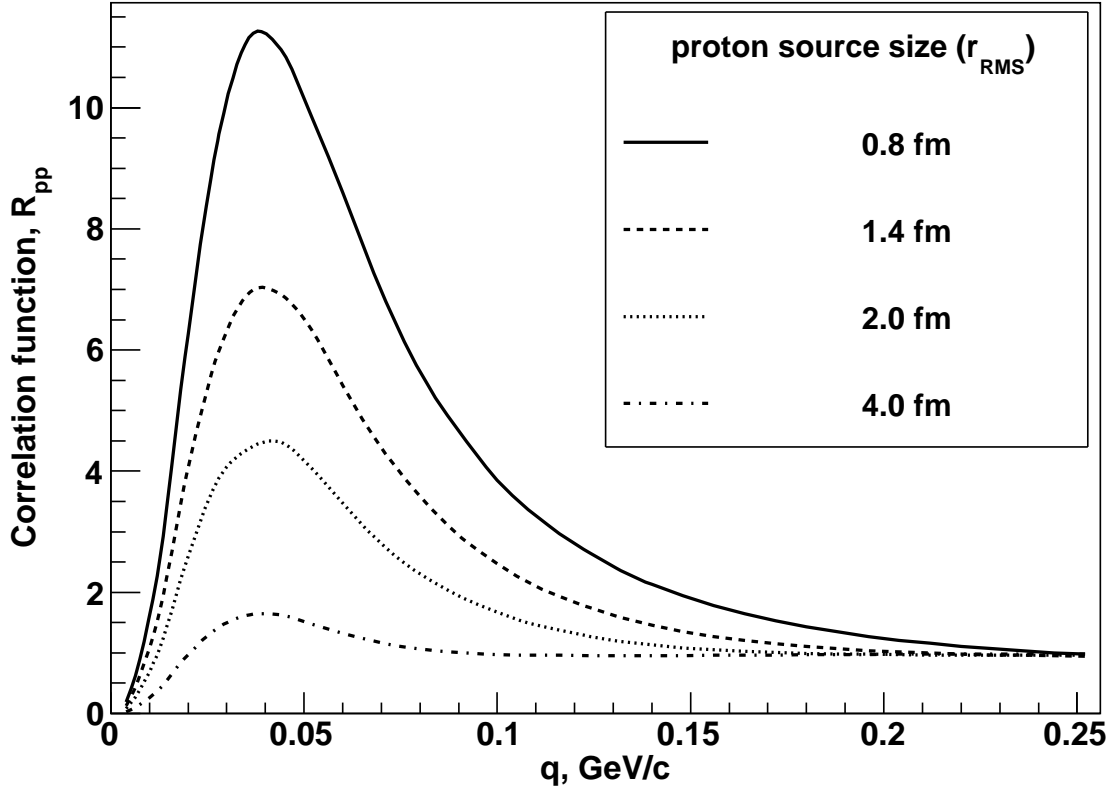}
\caption{The two-proton correlation functions $R_{pp}(q)$ are calculated 
for different source size parameters ($r_{RMS}$). Solid curve corresponds to $r_{RMS}=0.8$~fm.
Dashed curve corresponds to $r_{RMS}=1.4$~fm. Dotted curve  corresponds to $r_{RMS}=2.0$~fm.
Dash-dotted corresponds to $r_{RMS}=4.0$~fm.}
\label{fig7}
\end{figure}

Dividing secondary baryons into  groups: participants of dense cluster
which was formed in the  flucton-flucton interaction ($N_c$)  
and other participants ($N_p$)  (spectators are  not  in the  consideration)  
one can  expect hierarchy  of sizes $r(N_{c1},N_{c2})  < r(N_p,N_p )< $ $ < r(N_c ,N_p
)$. Such  measurements is  needed  to  control  the density  and  the
lifetime of  the baryonic system.   The example of two  proton correlation
functions  ($R_{pp}(q)$,  where  $\vec{q}  = \vec{p_1}  -  \vec{p_2}$,
$\vec{p_1}$ and  $\vec{p_2}$ are the individual proton  momenta in the
pair rest  frame) calculated for  expected source size values  ($r_{RMS}$ is
the root means square of the  source which the protons are emitted) is
shown in Fig.~\ref{fig7}.  The interference of identical particles, as
well as  Coulomb and strong final-state interactions,  were taken into
account.   Strong final-state  interactions are  dominant,  causing an
increase  of   the  pair  production   cross  section  near   $q  \sim
$~0.04~GeV/c. The  intensity of this  effect depends inversely  on the
source size parameter $r_{RMS}$.

Dense fermion-rich system should cause a decrease of the average distances
between  nonidentical baryons  in comparison  with identical  ones. 
Such  effect  is  interesting  itself  (see subsection  3.7  below).
Secondary  particle  momentum-space  region  where  correlations  
propose to study corresponds to the B2R3.  The pp(np) correlations should be measured
at relative momenta $q_0 < 0.2 GeV/c$ with relative momentum resolution $\Delta q \ll q_0$.

\subsection*{\hspace  {5  mm}\LARGE 3.3. Isosymmetrisation}

\hskip 0.7cm If dense  fermion rich  system is created  and selected  with trigger
(event selection)  and this fact is  ultimate (see section 3.1 and 3.2 above) one can
proceed with  its properties  study.  
All degrees  of freedom  in  such system which populating is not much energy 
consuming should be brought into play. 
Therefore  a broken symmetries  should be tend to
restore.   
Particularly, the  cross  section   ratio   of  particle
production when particle's are components of the same isomultiplet
should be close to unit~\cite{12}. 
This  conclusion is trivial for isosymmetric nuclei
collision  and  isosymmetric trigger, but  it becomes  non-trivial for  He$^3$+He$^3$
collisions and (or)  asymmetric trigger  (\textit{e.g.}  charged pion  or kaon).
The  proposed  measurements  are   $p/n$,  $\pi^+  /  \pi^-$  ratios  for
particles  produced in  selected  flucton-flucton interactions  within
B2R3  and  for  particles  outside  this kinematical  region  as  the
reference measurements. 
While  $\pi^+  /  \pi^-$ ratio measurement is a  routine task for tracking detector,
the   $n/p$ ratio measurement
needs special efforts. 
But ratio  measurement for nucleons seems to be
more  informative.  The  ``background''  measurements of  isosymmetric
nuclei  with   isosymmetric  trigger   are  needed  to   increase  the
measurement  precision of  isosymmetrisation effects  and  decrease of
systematical   errors.   Also   at   high  secondary particle multiplicity
isosymmetrisation  is trivial therefore  total multiplicity  should be
controlled.

\subsection*{\hspace  {5  mm}\LARGE 3.4. Strangeness}

\hskip 0.7cm Another broken  symmetry (SU(3)) also  should be tended to  restore in
high baryon  density conditions. This could cause  the equalization of
probabilities to produce different components of baryon octet. Since a
strange baryon (\textit{e.g.}  $\Lambda$-baryon) production must be accompanied
by production of additional  kaon (strangeness conservation), it will
result in noticeably  increase produced mass in the  process with free
energy lack.  At colliding nuclei energies  of a few GeV/nucleon
the energy  lack could be  regulated by varying of  minimal target
mass -- cumulative number  \cite{13}.  
By separating mass
increasing effect  from other  one  the probability  of strangeness  production
within  dense  baryonic system  would  be  higher  than the probability
of  non-strange particle production.   
Strangeness  increase is  also  considered  as  one of  the
signatures  of usual  quark-gluon plasma,  but the  reason for  that is
quite different (processes like $gg \rightarrow ss$)~\cite{14}.

\subsection*{\hspace  {5  mm}\LARGE 3.5. Vector mesons}  

\hskip 0.7cm An  increasing resonance production    and  high  spin particles  is
expected in cumulative processes.  In particular the vector to scalar
meson ratio  should increase with increasing of cumulative  number  (the effect
is predicted in~\cite{15}). An  interesting effects can be seen due to
free energy  lack  in the process.   Since kinematical restrictions
become  more  important with increasing of  the  produced  particle invariant  mass,
the  shape  and  width  of  peaks  which correspond  to  wide
resonances production  (like $\rho$ and  $\Delta$) are expected  to be
distorted with respect to PDG values.

\subsection*{\hspace  {5  mm}\LARGE 3.6. Exotic}

\hskip 0.7cm When the possibilities to  satisfy the requirements of Pauli principle
using known  degrees of  freedom are exhausted  then the  dense baryon
reach system  has to find new  forms of existence. The  role of exotic
states is expected to be increased in comparison with usual reactions in
dense fermionic  medium conditions. In particular  diquark medium will
help in  dibaryonic resonances  production. Exotic states  produced in
this processes  cannot be too  heavy due to kinematical  limits. Light
(below the  threshold with pions production)  pentaquarks like ($qqqqs$)
or (and) dibaryons  like ($qqqqqqqs$)  will probably decay  into nucleons
and photons.   Existing limits  of exotic production    \cite{16}
are  to several  orders magnitude  higher  than cross  section of  our
proposed  trigger hence  there is  no experimental  exclusion  for the
exotics discussed in the subsection.

\subsection*{\hspace  {5  mm} \LARGE 3.7. Multifermion effects}

\hskip 0.7cm Multifermion effects  could appear in  dense medium. To get  the first
imagination  on the  expected effects  one can  take into  account the
effects for dense multiboson  system, which was discussed, for example
in  \cite{17}.  While  for rare  systems  the slope  parameter of  the
momentum  spectra  and  the  size  parameter  for  the  width  of  the
interference effect  could be  independent, this is  not the  case for
dense matter. Equally populated cells in momentum and coordinate space
can be considered as an additional signature of dense matter.

\section*{\hspace  {5  mm}\LARGE 4. Concluding remarks}
\hskip 0.7cm        Some points  proposed program  already realized or  have status
``in    progress''    within    FLINT    experimental    program    at
ITEP~\cite{10}. The  whole proposed program  can be realized  in future
facilities FAIR (even with  SIS100) and (or) NICA (Nuclotron M).  Authors
would like to thank our  colleagues and collaborators from ITEP FLINT,
NICA  MPD and  FAIR  CBM and  especially  thanks to A.~A.~Baldin,  M.~Chernodub,
A.~Kaidalov,   D.~Kharzeev,    R.~Lednicky,   A.~Litvinenko,   L.~Malinina,
V.~Nikitin,  M.~Polikarpov, Yu.~Simonov,  S.~Shimansky~and M.~Tokarev  for
helpful  discussions on  related subjects.   This work  was  done with
financial  support  by Federal  agency  of  Russia  for atomic  energy
(Rosatom)    and    RFBF   under    grants    N   08-02-00676-a    and
08-02-92496-NCNIL-a.


\newpage

\end{document}